# Low-Contrast BIC Metasurfaces with Quality Factors Exceeding 100,000

Keisuke Watanabe[1]*, Tadaaki Nagao[1,2], and Masanobu Iwanaga[3]

1.* International Center for Materials Nanoarchitectonics (MANA), National Institute for Materials Science (NIMS), 1-1 Namiki, Tsukuba, Ibaraki 305-0044, Japan.
Email: watanabe.keisuke@nims.go.jp. https://orcid.org/0000-0002-4285-2135
2. Department of Condensed Matter Physics, Graduate School of Science, Hokkaido University, Kita 10, Nishi 8, Kita-ku, Sapporo 060-0810, Japan.
3. Research Center for Electronic and Optical Materials, National Institute for Materials Science (NIMS), 1-1 Namiki, Tsukuba, Ibaraki 305-0044, Japan.

**ABSTRACT:** Dielectric metasurfaces operating at quasi-bound states in the continuum (qBICs) can achieve exceptionally high radiative quality ($Q$) factors by introducing small asymmetries into their unit cells. However, fabrication imperfections often impose major limitations on the experimentally observed $Q$ factors. In this study, we experimentally demonstrate BIC metasurfaces with a record-high $Q$ factor of 101,486 under normal excitation of light in the telecom wavelength range achieved by employing low-contrast silicon pairs. Our findings show that such ultrahigh-$Q$ factors can be attained by leveraging both the high radiative $Q$ factors of higher-order qBIC modes and reduced scattering losses in shallow-etched designs. Additionally, we demonstrate stable sub-picometer-level wavelength fluctuations in water, with a limit of detection of $10^{-5}$ for environmental refractive index changes. The proposed approach can be extended to BIC metasurfaces with many other configurations and operating wavelengths for ultrahigh-$Q$ applications in both fundamental physics and advanced devices.

Strong confinement and localization of light in nanofabricated structures are of great importance for diverse applications, including high-efficiency lasers,[1] sensors,[2] bioimaging,[3] nonlinear enhancement,[4] non-Hermitian optics,[5] and topological photonics.[6] The energy dissipation of confined light is quantified by the quality ($Q$) factor, and both the material selection and structural design play major roles for maximizing the experimental $Q$ factors. To date, nanostructures with periodic arrays have been successfully employed to achieve high $Q$ factors. Such nanostructure arrays include photonic crystals (PCs),[7] plasmonic cavities,[8] and metallic or dielectric metasurfaces that support surface lattice resonances[9,10] and bound states in the continuum (BICs)[11]. Remarkably, BICs in dielectric materials have recently attracted considerable attention owing to their design flexibility in controlling radiative losses through precise structural design. A symmetry constraint dictates that symmetry-protected BICs are not accessible from free-space. However, true BICs can be transformed into quasi-BICs (qBICs) by breaking the symmetry of the unit cell, yielding finite radiative $Q$ factors. This symmetry breaking enables the observation of sharp resonances under normal-incidence excitation. The high radiative $Q$ factors and resonantly enhanced electric fields are the key features of the qBIC modes, offering novel strategies to enhance the functionality of optical devices.[12–14] However, BIC metasurfaces, which often have a large ratio of nanostructure depth to in-plane dimensions, are susceptible to scattering losses owing to fabrication imperfections, thus limiting their experimental $Q$ factors to the range of several hundreds to thousands. More specifically, the localized electric fields within high-index nanostructures overlap with the sidewalls, on which light scattering caused by surface roughness from nanofabrication drastically reduces the experimental $Q$ factors. In other words, experimental $Q$ factors cannot improve under large scattering losses, despite many researchers focusing on increasing the radiative $Q$ factors obtained from numerical calculations. To resolve this problem, two approaches can be considered. The first approach involves utilizing imperfection-tolerant designs.[15,16] Jin et al. proposed that merging off-Γ BICs with multiple topological charges into an isolated symmetry-protected BIC at the Γ point can preserve high $Q$ factors over a broad range in the $k$-space, substantially suppressing out-of-plane scattering.[15] This approach enabled the experimental demonstration of ultrahigh-$Q$ factors up to $4.9 \times 10^5$ in PC structures. However, topological charge engineering requires precise control in fabrication, because the merging BICs are extremely sensitive to structural parameters.[17] The second approach focuses on imperfection-reduced designs. Huang et al. recently demonstrated a simple structure with a thin patterned photoresist layer on top of a silicon-on-insulator (SOI) wafer,[18] achieving an ultrahigh-$Q$ guided mode resonance with a $Q$ factor as high as $2.4 \times 10^5$ for PC structures. In this design, the absence of nanopatterning in the silicon layer minimized scattering losses typically caused by surface roughness from silicon etching. However, this ultrahigh-$Q$ factor has not been demonstrated in BIC metasurfaces to date. More importantly, nanopatterned photoresist layers suffer from poor durability because they can be easily damaged, degraded, and dissolved in many solvents. Therefore, achieving monolithic high-$Q$ metasurfaces with stable patterned layers remains challenging, which is essential for a wide range of optical device applications.

In this study, we propose and experimentally demonstrate a method to minimize fabrication imperfections in silicon metasurfaces by employing shallow-etched structures.[19–21] Our approach increases the experimental $Q$ factors by approximately one or two orders of magnitude compared with conventional designs, pushing the limits of $Q$ factors over $10^5$. We design, fabricate, and characterize BIC metasurfaces composed of arrays of silicon pairs with varying etching depths. Our results show that shallower etching simultaneously enhances the radiative $Q$ factors and reduces scattering losses from fabrication imperfections in higher-order qBIC modes, leading to substantially improved experimental $Q$ factors. Finally, we demonstrate highly stable refractometric sensing, achieving a limit of detection (LOD) at the $10^{-5}$ level using low-contrast BIC metasurfaces. We expect that the developed ultrahigh-$Q$ silicon metasurfaces will find broad applications in fields requiring strong light–matter coupling at the nanoscale.

Figure 1a illustrates the proposed nanostructures, which are shallow-etched silicon pairs with etching depth $d$, fabricated on SOI wafers with a thickness of 400 nm. The buried oxide (BOX) layer is 2000 nm, which is thick enough to sufficiently suppress leakage losses to the bottom silicon substrate.[22] The dimensions of the unit structure are period $P$ = 760 nm, shallow rod length $L$ = 610.8 nm, and shallow rod width $w$ = 235.5 nm. The symmetry of the unit structure is broken by changing asymmetry parameter $\alpha = 2\Delta L/L$, which controls the radiation losses of the resonance mode emitted into the far-field. Figure 1b shows the simulated transmittance spectra for an infinite periodic structure computed using the finite-difference time-domain (FDTD) method (Ansys Lumerical). The resonance modes are excited by a normally incident $x$-polarized plane wave parallel to the major axis of the rectangular structures. For $d$ = 0 (i.e., unpatterned silicon), the transmittance spectrum exhibits interference patterns owing to the multi-layer configuration of the SOI wafer. When shallow-etched nanostructures with $d$ = 82.7 nm and $\alpha$ = 0% (i.e., no asymmetry) are introduced, a large transmittance dip appears around the wavelength of 1870 nm caused by the destructive interference between the leaky guided-mode resonance and background continuum (Figure S2). When $\alpha$ = 5%, three BIC modes are transformed into qBIC modes. The cross-sectional $E_x$ profiles shown in Figure 1c indicate that the metasurfaces support a fundamental mode (qBIC1) and higher-order modes (qBIC2 and qBIC3) in the wavelength range of interest. Figure 1d (left) presents the transmittance maps for different asymmetries $\alpha$ when $d$ = 82.7 nm. The three qBIC modes retain narrow linewidths even as $\alpha$ increases, suggesting that the low-contrast matesurfaces maintain large radiative $Q$ factors across a broad range of $\alpha$. Notably, the peak wavelengths of the three qBIC modes remain nearly constant as $\alpha$ changes because the overall volume of the shallow pair-rod remains unchanged despite changes in the lengths of the upper and lower rods. Figure 1d (right) shows the transmittance maps for different $d$ when $\alpha$ = 5%. As $d$ increases, the linewidths gradually broaden, and the resonance wavelengths undergo a blueshift, as the modes penetrate more into surrounding air. In the following, we focus on the qBIC2 mode to achieve ultrahigh-$Q$ factors because the localized electric field within the silicon layer is less affected by $d$ for the qBIC2 mode than by $d$ for the qBIC1 and qBIC3 modes, as evidenced by the smaller slope of the resonance wavelength with respect to $d$ (Figure 1d). In other words, the qBIC2 mode is intrinsically well-confined within the silicon layer and less sensitive to the sidewalls in the shallow pair-rod nanostructures.

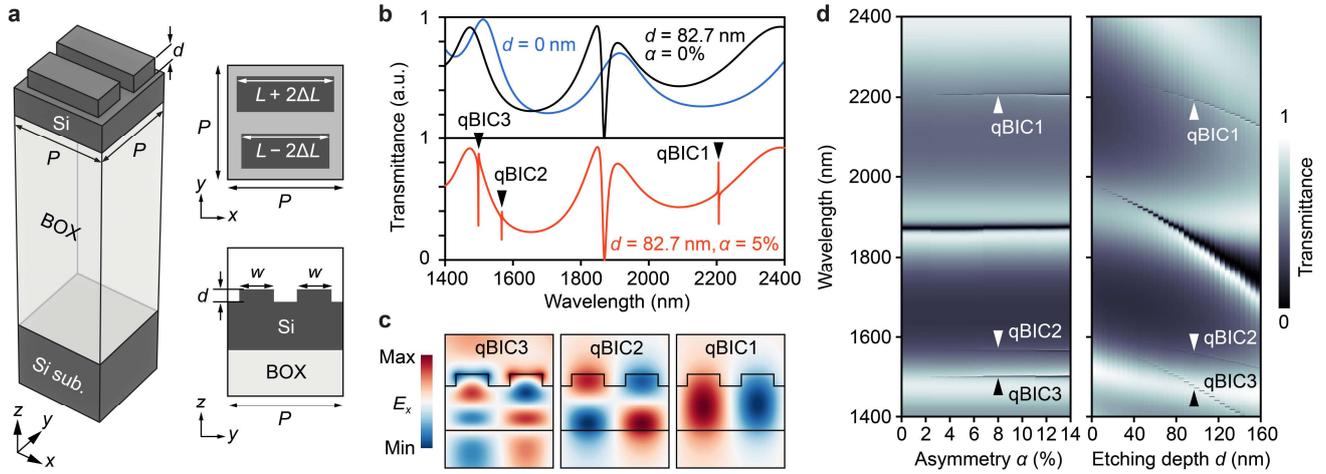

Figure 1. Low-contrast BIC metasurfaces fabricated on SOI wafers. (a) Schematic of structure and its dimensions. (b) Simulated spectra for unpatterned silicon (blue) and metasurfaces with $d$ = 82.7 nm for $\alpha$ = 0% (black) and $\alpha$ = 5% (red). (c) Cross-sectional $E_x$ distributions for three qBIC modes supported by proposed metasurfaces with $d$ = 82.7 nm and $\alpha$ = 5%. (d) Simulated transmittance maps for different $\alpha$ when $d$ = 82.7 nm (left) and for different $d$ when $\alpha$ = 5% (right).

For the experimental realization of the designed low-contrast metasurfaces, electron beam lithography with a positive resist was employed, followed by dry etching using the Bosch process with $SF_6$ and $C_4H_8$ gases, allowing the precise control of the silicon etching depth (Supporting Information S1). Figure 2a, b presents scanning electron microscopy (SEM) images of the fabricated metasurfaces under highly controlled etching depth conditions. The fabricated device was characterized using a custom-built setup comprising a tunable laser and photodiode (Supporting Information S1). Figure 2c shows the simulated and experimental transmittance spectra for $d$ = 82.7 nm when normally incident $x$-polarized light was coupled with the metasurfaces with different $\alpha$. The simulations and experiments were in good agreement, showing the qBIC2 resonance mode even for a small asymmetry $\alpha$ of 0.5%. Because the length difference between the upper and lower shallow rods for $\alpha$ = 0.5% is given by $(L + 2\Delta L) - (L - 2\Delta L)$ = 6.1 nm, the fabrication disorder is considered to be smaller than this value. As seen in the spectra, the resonance amplitude increased with increasing $\alpha$, which is a typical behavior of qBIC modes. However, small sidebands occasionally appeared in the spectrum, possibly due to variations in size of the fabricated structures or interference effects caused by the large-area nanostructure arrays. Additionally, a slight blueshift in the experimental resonance wavelengths compared with the simulations was observed, which can be attributed to the rounded corners from lithography and/or oblique sidewalls from etching. Figure 2d shows a representative enlarged SEM image and the corresponding spectrum for $\alpha$ = 1%, showing a clear ultrasharp Fano resonance. The $Q$ factor was extracted by fitting the transmittance spectrum with a Fano function, yielding a record-high $Q$ factor of 101,486 at resonance peak wavelength of $\lambda$ = 1560.3 nm.

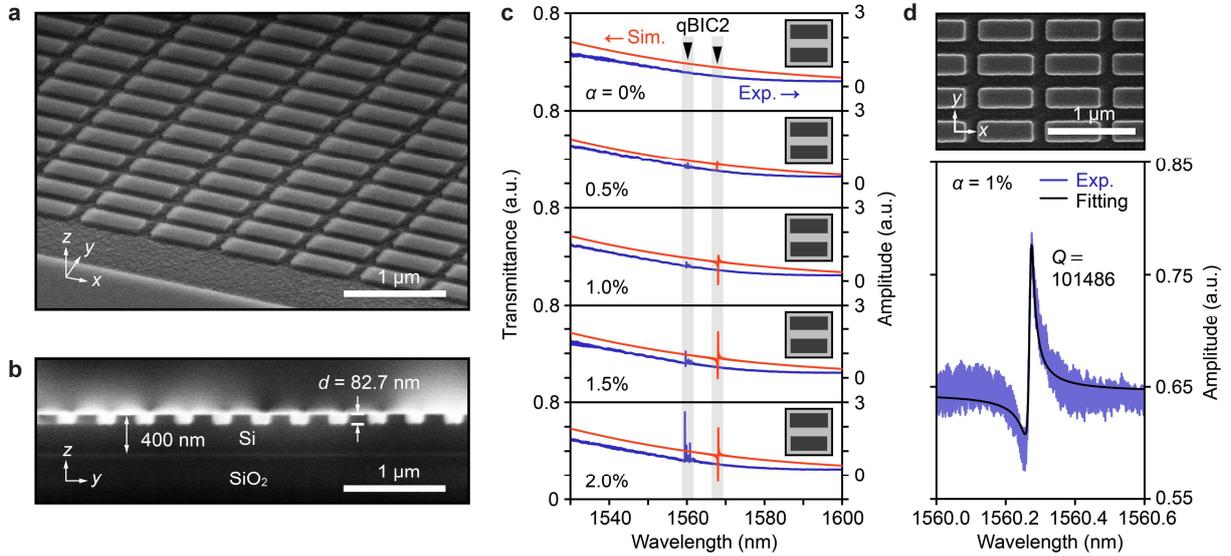

Figure 2. Fabricated low-contrast BIC metasurfaces with $d$ = 82.7 nm. (a) Tilted and (b) $yz$-plane cross-sectional SEM images. (c) Simulated (red) and experimental (blue) transmittance spectra for different $\alpha$. The gray regions indicate the wavelengths where the qBIC2 mode appears. Each inset shows a schematic of the unit structure with broken symmetry. (d) Representative SEM image (top) and transmittance spectrum (bottom) of metasurface with $\alpha$ = 1%. The $Q$ factor is extracted by fitting the transmittance spectrum with a Fano function.

To further elucidate the physics underlying the ultrahigh-$Q$ factors, low-contrast BIC metasurfaces with etching depths $d$ of 82.7 nm, 116.1 nm, and 149.5 nm were fabricated, and their $Q$-analysis was conducted by varying asymmetry $\alpha$. First, we compared the simulated transmittance spectra for the metasurfaces with three etching depths, as shown in Figure 3a. As $\alpha$ increased, the resonance peaks blueshifted, and the resonance linewidths broadened, corresponding to an increased radiative component coupling into the external medium. The experimentally measured $Q$ factors as a function of $\alpha$ are shown in Figure 3b. The black curves overlapping the experimental $Q$ factors were fitted using $Q^{-1} = Q_r^{-1} + Q_{scat}^{-1}$, where $Q_{scat}$ was determined using nonlinear least squares curve-fitting method and assuming that $Q_{scat}$ was independent of $\alpha$. $Q_{scat}^{-1}$ represents the scattering losses arising from fabrication imperfections. The loss component that does not contribute to far-field coupling is expressed as the sum of the material absorption and scattering losses owing to fabrication imperfections. However, in this study, we only considered scattering losses because the silicon and BOX layers exhibit negligible absorption in the wavelength range of interest. $Q_r$ denotes the radiative $Q$ factor calculated from complex eigenfrequencies obtained from the finite element method (FEM) method (COMSOL). Here, $Q_r$ follows the typical inverse square relation, $Q_r = Q_0 \alpha^{-2}$ for qBIC modes,[23] with $Q_0$ being a constant that depends on the metasurface design and mode radiation characteristics (Figure 3c). As shown in Figure 3b, for small $\alpha$, the experimental $Q$ factors largely deviated from the inverse square relation for all etching depths and approached a fix value determined by $Q_{scat}$. These behaviors indicated that the experimental $Q$ factors were limited by an inherent $Q_{scat}$. The difference between the experimental $Q$ factors and $Q_r$ widened as $d$ increased, indicating an increase in scattering losses (i.e., decrease in $Q_{scat}$) for larger etching depths. Figure 3d shows the experimentally determined $Q_{scat}$ as a function of $d$, demonstrating that metasurfaces with smaller $d$ exhibited reduced scattering losses, thereby yielding higher experimental $Q$ factors. Although the highest $Q$ factor exceeding $10^5$ was achieved for $d$ = 82.7 nm in our experiments, further reductions in $d$ below 50 nm did not yield any additional increase in $Q_{scat}$ (Figure S3). This suggests that once a certain $Q_{scat}$ is reached, further improvements in $Q$ factors are unlikely due to inevitable fabrication errors and structural disorder.

As discussed below, the condition where $Q_r = Q_{scat}$ corresponds to the critical coupling condition, which plays a crucial role in various applications such as sensing.[24–26] In our experiment, we selected and measured metasurfaces with $\alpha$ = 3%, 4%, and 5% for etching depths $d$ = 82.7 nm, 116.1 nm, and 149.5 nm, respectively, as these structural conditions were close to the critical coupling condition ($\alpha_{cc}$). The representative spectra are shown in Figure 3e. As expected, the experimental $Q$ factors increased as $d$ decreased. Figure 3e also shows the cross-sectional electric field distributions, indicating that the overlap between the localized electric fields and sidewalls of the shallow pair-rod structures reduced with decreasing $d$. This observation supports our conclusion that the increased $Q$ factors for smaller $d$ are attributable to the reduced ratio of nanostructure depth to in-plane dimensions, resulting in lower scattering losses due to nanostructured sidewall roughness. Considering that $Q_0$ is larger for smaller $d$, we can conclude that the experimental $Q$ factors increased by both the high $Q_r$ and $Q_{scat}$ values for the low-contrast metasurfaces.

Table 1 compares the ultrahigh-$Q$ factors achieved in this study with previously reported experimental $Q$ factors (exceeding 1000) for all-dielectric metasurfaces. The experimental $Q$ factor of our low-contrast BIC metasurface was one order of magnitude higher than the highest $Q$ factor of 18,511 reported to date.[27] Given that most reported $Q$ factors were 1000 or less,[14,28–30] the $Q$ factor of our metasurface was one or two orders of magnitude higher than typical values.

Table 1. Experimentally observed *Q* factors above 1000 reported to date in all-dielectric metasurfaces.

| Unit structure | Resonance type | Material | Peak λ (nm) | Exp. *Q* factor | Ref. |
|---|---|---|---|---|---|
| Cylinder | SP-BIC | SOI | 1425 | 1946 | [31] |
| T-shape block | SP-BIC | Si on quartz | 1588 | 18,511 | [27] |
| U-shape block | SP-BIC | Si on sapphire | 1548 | 3534 | [32] |
| Tilted bars | Chiral-BIC | $TiO_2$ on $SiO_2$ | 612 | 1250 | [33] |
| Cuboid | Accidental-BIC | SOI | 1538 | 5305 | [34] |
| Block with nanogaps | SP-BIC | Si on quartz | 1553 | 1233 | [35] |
| Shallow tetramer | SP-BIC | $Si_3N_4$ on quartz | 828 | 6061 | [21] |
| Square nanodisk with hole | TD-BIC | SOI | 1497 | 4990 | [36] |
| Nanodisk dimer | TD-BIC | SOI | 1480 | 3142 | [37] |
| Nanorod | SP-BIC | a-Si on fused silica | 1505 | 4130 | [38] |
| Nanodisk | SLR | a-Si on silica | 1183 | 2750 | [39] |
| Square nanopillar | SP-BIC | SOI | 1685 | 2476 | [40] |
| **Shallow pair-rod** | **SP-BIC** | **SOI** | **1560** | **101,486** | **This study** |

SP, symmetry-protected; TD, troidal dipole; a-Si, amorphous silicon; SLR, surface lattice resonance

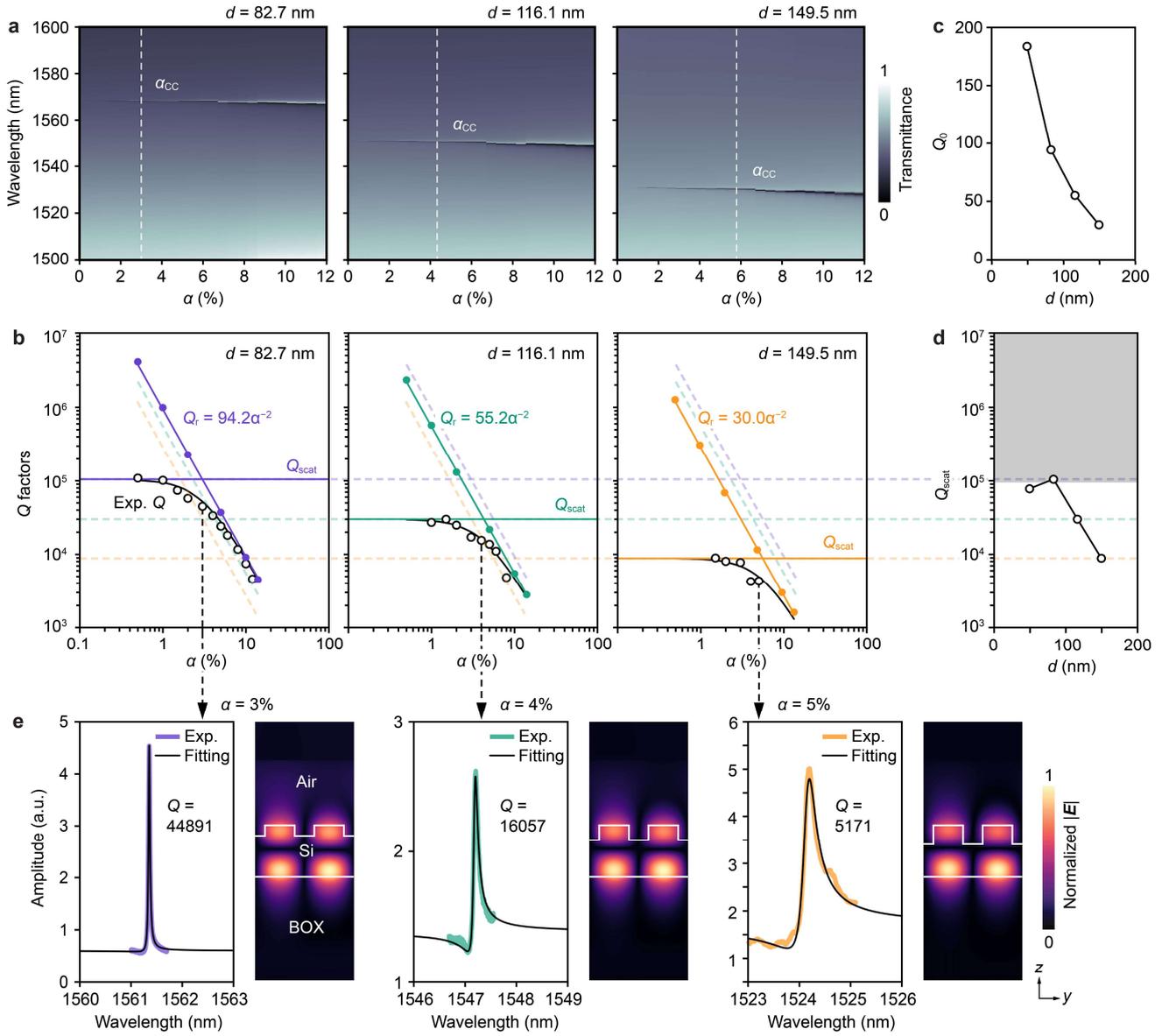

Figure 3. Characteristics of low-contrast BIC metasurfaces with different etching depths $d$. (a) Transmittance maps. The dashed lines indicate $\alpha$ values that satisfy the critical coupling conditions ($\alpha_{CC}$). (b) $Q$-analysis results. The radiative $Q$ factors ($Q_r$, filled circles) with their fitted lines and $Q_{scat}$ obtained by fitting the experimental $Q$ factors (black open circles) are shown. The $Q$ factors were measured in air. (c) Coefficient $Q_0$ obtained by fitting inverse square relation of $Q_r$. (d) Extracted $Q_{scat}$. The gray region indicates the approximate upper limit of $Q_{scat}$. (e) Experimental spectra and corresponding Fano fitting for metasurfaces near critical coupling conditions: $\alpha = 3\%$ for $d = 82.7$ nm, $\alpha = 4\%$ for $d = 116.1$ nm, and $\alpha = 5\%$ for $d = 149.5$ nm. The normalized electric field $|E|$ distributions for each condition are also shown.

Next, we characterized the sensing properties of the low-contrast BIC metasurfaces. Previous studies have reported that the lowest LOD for environmental refractive index changes can be achieved under critical coupling conditions.[26,41] Therefore, we compared the refractometric sensing performance of metasurfaces near these critical coupling conditions for different etching depths $d$. Figure 4a shows the experimental results, where aqueous solutions with different bulk refractive indices, adjusted by mixing isopropyl alcohol (IPA) and heavy water ($D_2O$), were introduced into a polydimethylsiloxane (PDMS) microfluidic channel. The resonance peak wavelengths were recorded in real-time. Here, we used $D_2O$ instead of $H_2O$ to avoid absorption loss of water in the wavelength range of interest, thus simplifying the analysis of sensing performance. As shown in the figure, the resonance peak wavelengths redshifted with increasing refractive index of the solution. Figure 4b shows the relation between the resonance peak wavelength shift $\Delta\lambda$ and refractive index change $\Delta n$. The environmental refractive index sensitivity $S$ was calculated from the slope and found to be approximately 26 nm/RIU, which was smaller than the typical $S$ of several hundred of nm/RIU for photonic sensors.[42–45] Although $S$ increased slightly with increasing $d$, the increment was small, being consistent with the simulation results (Figure S4). This behavior is also attributed to the intrinsically strong confinement of the higher-order qBIC2 mode used in this study. Figure 4c shows real-time measurements of peak wavelength fluctuations $\delta\lambda$ (= $3\sigma$, where $\sigma$ is the standard deviation) in $D_2O$ over a period of approximately 1 min. The $\delta\lambda$ increased slightly as linewidths widened with increasing $d$. Here, the short evaluation period (approximately 1 min) was chosen to minimize the effects of wavelength drift. The minimum $\delta\lambda$ was 0.79 pm for the metasurface with $d$ = 82.7 nm ($\alpha$ = 3%). Figure 4d-g compares the experimental $Q$ factors, sensitivity $S$, figure-of-merit (FOM), and LOD for different $d$. The FOM defined as FOM = $S$/FWHM, where FWHM is the full width at half maximum, reached a maximum of 829 when $d$ = 82.7 nm (Figure 4f). This FOM was among the highest reported for similar all-dielectric metasurfaces. The LOD given by $\delta\lambda/S$ slightly reduced with decreasing $d$ (Figure 4e), yielding values of $3.00 \times 10^{-5}$, $3.28 \times 10^{-5}$, and $3.51 \times 10^{-5}$ for $d$ = 82.7 nm, 116.1 nm, and 149.5 nm, respectively. Although the wavelength fluctuations were small at the sub-picometer scale, sensitivity $S$ was also small. Therefore, increasing the $Q$ factors does not lead to substantial LOD improvements due to the general trade-off between the $Q$ factor and sensitivity. Nevertheless, our ultrahigh-$Q$ silicon metasurfaces hold great potential for applications that require detection of local perturbations, such as single-molecule detection at high concentrations, leveraging both their high $Q$ factors and small mode volume.[46] Moreover, the proposed metasurfaces offer substantial advantages, including simple measurements based on position-insensitive vertical excitation from free-space and availability of well-established complementary metal–oxide–semiconductor compatible fabrication processes, facilitating the practical implementation of low-cost sensing systems.

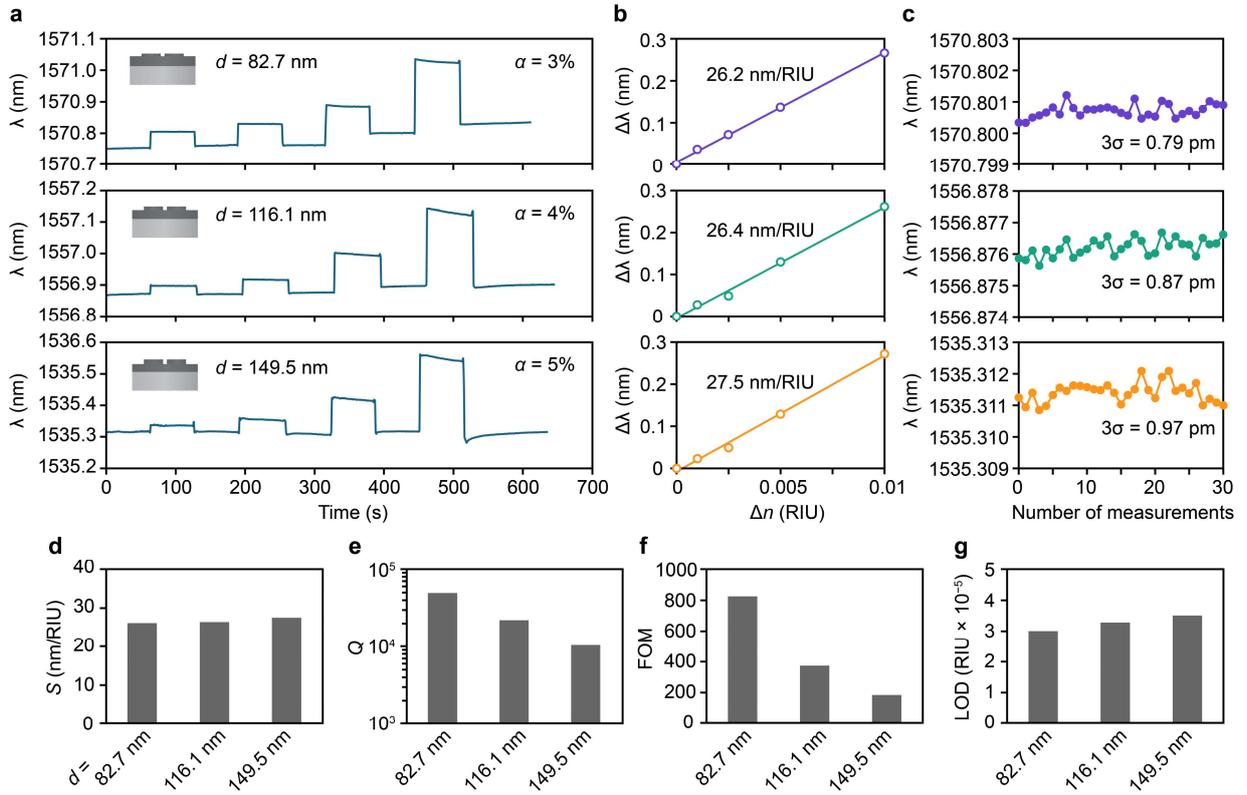

Figure 4. Refractometric sensing in low-contrast BIC metasurfaces with varying etching depths $d$. (a) Real-time measurement results. $D_2O$ solutions with different refractive indices were introduced sequentially, with refractive index variations of $\Delta n$ = 0.001, 0.0025, 0.005, and 0.01. After each step, the metasurfaces were rinsed with $D_2O$. (b) Wavelength shift as a function of refractive index variation. Sensitivity $S$ was derived from the curve slope. (c) Representative real-time fluctuations in resonance peak wavelength measured in $D_2O$. (d)-(g) Comparison of $S$, $Q$, FOM, and LOD for different $d$.

In conclusion, we have experimentally demonstrated ultrahigh-$Q$ factors exceeding $10^5$ in low-contrast silicon metasurfaces supporting higher-order qBIC modes. By designing the metasurfaces to minimize the overlap between the localized electric fields and sidewalls in shallow-etched nanostructures, we achieved high radiative $Q$ factors while reducing scattering losses from fabrication imperfections. Specifically, for an etching depth of 82.7 nm and asymmetry parameter of 1%, we obtained a record-high $Q$ factor of 101,486, which was one or two orders of magnitude higher than that of typical BIC metasurfaces. Additionally, we observed sub-picometer peak wavelength fluctuations in an aqueous solution, demonstrating an improved limit of detection in the order of $10^{-5}$ for changes in the environmental refractive index. Given that qBIC modes can be easily coupled to normally incident light without the need for delicate coupling systems, we believe that our ultrahigh-*Q* metasurfaces offer a promising platform for a wide range of applications requiring strong light–matter coupling.

**Supporting Information**

Additional details about simulation, fabrication, and optical characterization; electric field distributions; *Q*-analysis for metasurfaces with shallower etching depth; simulation results of refractometric sensitivity (PDF)

**Acknowledgement**

This work was financially supported by JSPS KAKENHI Grant Number JP22K20496, Iketani Science and Technology Foundation (Grant Number 0361252-A), and "Advanced Research Infrastructure for Materials and Nanotechnology in Japan (ARIM)" of the Ministry of Education, Culture, Sports, Science and Technology (MEXT). Proposal Number JPMXP1223NM5060, JPMXP1224NM5259.

**Conflict of interest**

The authors declare no competing interests.

# Supporting Information for

**Low-Contrast BIC Metasurfaces with Quality Factors Exceeding 100,000**

Keisuke Watanabe[1]*, Tadaaki Nagao[1,2], and Masanobu Iwanaga[3]

1.* International Center for Materials Nanoarchitectonics (MANA), National Institute for Materials Science (NIMS), 1-1 Namiki, Tsukuba, Ibaraki 305-0044, Japan.
2. Department of Condensed Matter Physics, Graduate School of Science, Hokkaido University, Kita 10, Nishi 8, Kita-ku, Sapporo 060-0810, Japan.
3. Research Center for Electronic and Optical Materials, National Institute for Materials Science (NIMS), 1-1 Namiki, Tsukuba, Ibaraki 305-0044, Japan.

* Email: watanabe.keisuke@nims.go.jp

## S1. Materials and methods

**Simulations**: Transmittance spectra and electromagnetic mode profiles were simulated using a commercial FDTD solver (Ansys Lumerical). Radiative $Q$ factors were calculated using the FEM (COMSOL Multiphysics) and determined using equation $Q = \text{Re}(f)/2\text{Im}(f)$, where $f$ is the complex eigenfrequency. In both simulations, perfectly matched layers were applied along the $z$ axis, while Bloch (periodic) boundary conditions were used along the $x$ and $y$ axes for the FDTD (FEM).

**Fabrication**: To fabricate low-contrast silicon metasurfaces, SOI wafers with a 400 nm top silicon layer and a 2000 nm BOX layer were initially cleaned in an ultrasonic bath with acetone and IPA, followed by $O_2$ plasma treatment (AQ-500, Samco). A 100 nm-thick positive resist (ZEP-520A, Zeon Chemicals) diluted 1:1 in anisole was spin-coated onto the SOI wafers and prebaked at 180°C for 3 min on a hotplate. A nanopattern with a sufficiently large dimension of 100 × 100 μm was defined via electron beam lithography (ELS-BODEN, Elionix) at an acceleration voltage of 100 kV. After patterning, the sample was developed in xylene for 1 min and rinsed in IPA for 30 s. The nanopatterns were then transferred into the silicon layer using a Bosch process via dry etching (MUC-21 ASE-SRE, Sumitomo Precision Products) based on $SF_6$ and $C_4H_8$ gases. The silicon etching depth was controlled by adjusting the number of etching and passivation cycles in the Bosch process. The etching rate was measured using a spectroscopic ellipsometer (M-2000, J.A.Woollam). Any remaining resist was finally removed by water vapor plasma treatment (AQ-500, Samco).

**Characterization**: The transmittance spectra were characterized using a custom-built setup (Figure S1). A tunable single-mode laser (TSL-510, santec) light with *x*-polarization was directed vertically onto the metasurface through a 10× objective (M Plan Apo NIR, NA = 0.26, Mitsutoyo), and transmitted light was detected by an InGaAs photodiode through a 5× objective (M Plan Apo NIR, NA = 0.14, Mitsutoyo). Background noise was eliminated using a cross-polarized configuration with two polarizers. The photodiode signal was acquired using a data acquisition (DAQ) device (USB-6212, NI) synchronized with the tunable laser (sweep speed of 1 nm/s) and recorded via a LabVIEW (NI) program at a 10 kHz sampling rate. For refractometric sensing, the fabricated metasurfaces were integrated into a custom-made PDMS microfluidic channel (Micro TAS Engineering), and the liquid solution was exchanged using a peristaltic pump (Takasago Fluidic Systems). The refractive index of the solution was pre-measured using a refractometer (PAL-RI, Atago) and adjusted by mixing deuterium oxide $D_2O$ (151882, Sigma-Aldrich) with IPA. During real-time measurements, the resonance peak wavelengths were determined by fitting the transmittance spectra with Fano profiles.

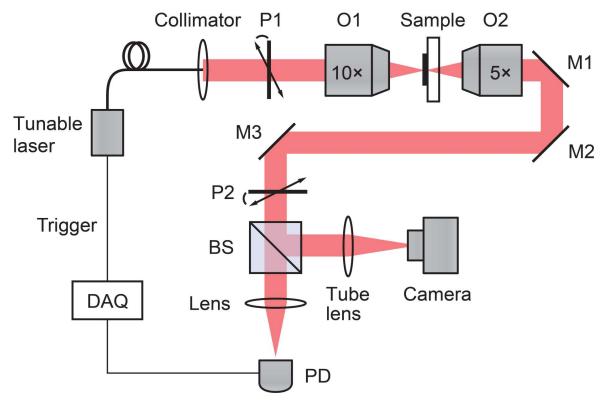

Figure S1. Experimental setup for transmittance measurement. P1 and P2, polarizers; O1 and O2, objectives; M1, M2, and M3, mirrors; BS, beam splitter; PD, photo detector.

**S2. Electric field distributions**

Figure S2 shows the classifications and mode profiles for the four modes discussed in the main text. The cross-sectional $E_x$ profiles are shown for the transverse electric (TE)-like qBIC1, qBIC2, guided-mode resonance (GMR), and qBIC3 modes on the $xz$ plane (at the center of the lower rod) and $yz$ plane (at the center of the unit cell). The qBIC2 mode exhibits small mode localization into surrounding air outside the silicon on both the $xz$ and $yz$ planes, which can be attributed to the high $Q$ factor of qBIC2.

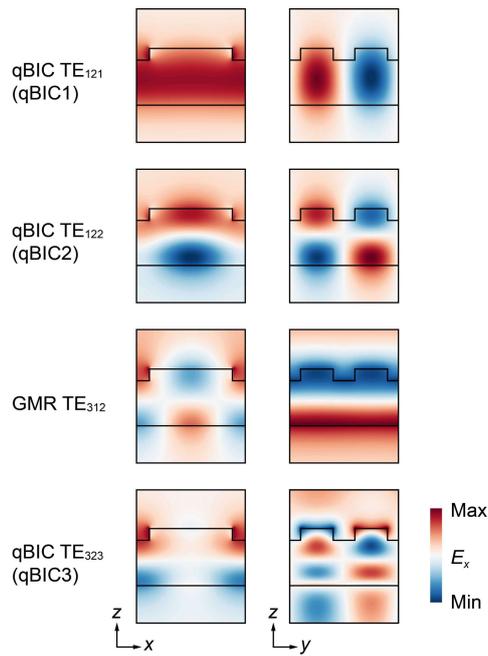

Figure S2. Cross-sectional $E_x$ distributions for qBIC1, qBIC2, GMR, and qBIC3 modes supported by metasurfaces with $d$ = 82.7 nm and $\alpha$ = 5%.

**S3. *Q*-analysis for metasurfaces with shallower etching depth**

In the main text, we mention that the experimental *Q* factors did not notably increase for etching depth *d* below 50 nm. To clarify this, Figure S3a,b shows the *Q*-analysis results for the metasurfaces with *d* = 49.3 nm and compares them to those for *d* = 82.7 nm. As expected, the radiative *Q* factor $Q_r$ increases for *d* = 49.3 nm. However, the experimental *Q* factors were comparable to those for *d* = 82.7 nm, particularly when *α* was small. This result can be explained by the scattering loss component ($Q_{scat}$) reaching a limit of approximately $10^5$, as shown in Figure S3c.

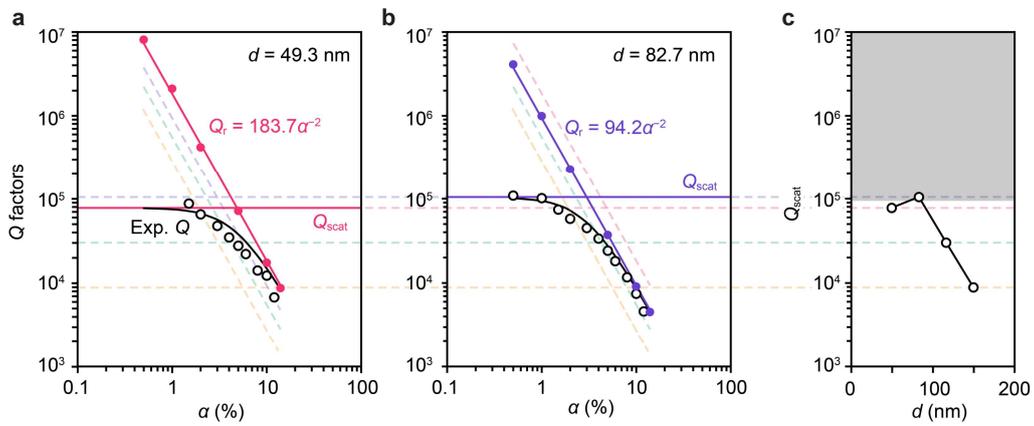

Figure S3. Comparison of *Q* factors for (a) *d* = 49.3 nm and (b) *d* = 82.7 nm. (c) Extracted $Q_{scat}$ as a function of etching depth *d*.

**S4. Simulation results of refractometric sensitivity**

The refractive index sensitivity of low-contrast BIC metasurfaces for different $\alpha$ and $d$ was calculated using the FEM by analyzing the shift in eigenfrequencies owing to changes in the environmental refractive index. We confirmed that $S$ increases slightly as $d$ increases (Figure S4), which can be attributed to the reduction in the mode confinement inside the silicon layer. Although the experimental $S$ (Figure 4b) was lower than the calculated values, the difference in $S$ between different $d$ was consistent with the simulation results.

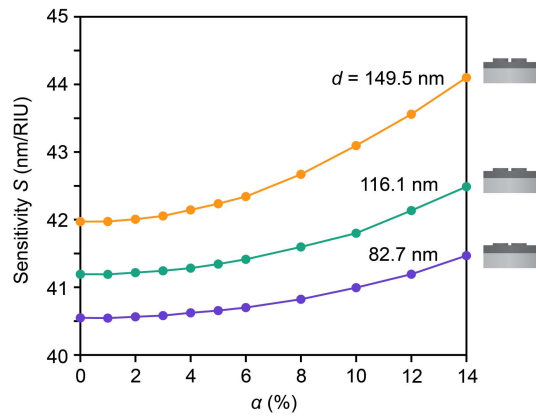

Figure S4. Calculated refractometric sensitivity $S$ as a function of asymmetry $\alpha$ for different $d$.